\documentclass{ws-ijmpcs}

\def\trimmarks{}

\newcommand{\be}{\begin{equation}}
\newcommand{\ee}{\end{equation}}
\newcommand{\bea}{\begin{eqnarray}}
\newcommand{\eea}{\end{eqnarray}}

\newcommand{\bfk}{\mbox{\boldmath $k$}}

\newcommand{\bfp}{\mbox{\boldmath $p$}}

\newcommand{\bfP}{\mbox{\boldmath $P$}}

\newcommand{\pup}{p^\uparrow}

\def\lsim{\mathrel{\rlap{\lower4pt\hbox{\hskip1pt$\sim$}}\raise1pt\hbox{$<$}}}
\def\gsim{\mathrel{\rlap{\lower4pt\hbox{\hskip1pt$\sim$}}\raise1pt\hbox{$>$}}}
\def\nostrocostruttino#1\over#2{\mathrel{\mathop{\kern 0pt \rlap
{\hbox{$#1$}}} \hbox{\kern-.135em $#2$}}}

\def\bt{b_T}

\def\kt{k_\perp}

\def\pp{p_\perp}

\def\avk{\langle k_\perp ^2\rangle}
\def\avp{\langle p_\perp ^2\rangle}

\begin{document}

\markboth{Mariaelena Boglione}
{Phenomenology of TMDs}

%
\catchline{}{}{}{}{}
%

\title{PHENOMENOLOGY OF TRANSVERSE MOMENTUM DEPENDENT PARTON DENSITIES
}

\author{MARIAELENA BOGLIONE
}

\address{Physics Department, University of Torino, \\Via P. Giuria 1
Torino, 10125, Italy
\\
boglione@to.infn.it}
%
%

\maketitle

\begin{history}
\received{Day Month Year}
\revised{Day Month Year}
\end{history}

\begin{abstract}
This talk provides a short overview of the phenomenology of transverse momentum 
depedent distribution and fragmention functions, focussing on the most recent 
phenomenological developments in the study of their $Q^2$ evolution and energy 
depedence. 

\end{abstract}

\ccode{PACS numbers: 13.88.+e, 12.38.Bx, 13.85.Ni}

The exploration of the three-dimensional structure of the nucleon,
both in momentum and in configuration space, is one of the
major issues in high energy hadron physics. 
Transverse momentum dependent parton densities (TMDs)
have recently attracted huge interest, 
not only for their remarkable spin correlation properties, 
but mostly because they represent a crucial tool for the investigation  
of the three-dimensional structure of the nucleon.

Huge amount of experimental data on spin asymmetries
in several different processes show that TMD distribution
and fragmentation functions exist and are non zero. 
Much progress has been achieved, for instance, in the phenomenological 
studies of the Sivers distribution function, which represents the number density  
of unpolarized partons inside a transversely polarized hadron, and of 
the Collins fragmentation function, which is related to the probability of a 
polarized quark fragmenting into an unpolarized hadron.

TMD studies are readily performed in the framework of QCD factorization, 
within a generalized parton model that incorporates the partonic intrinsic 
transverse motion in the kinematics of the examined scattering processes.
In this simple phenomenological approach, cross sections and
spin asymmetries are generated as convolutions of distribution 
and (or) fragmentation TMDs with elementary scattering cross sections; 
for instance for semi-inclusive deep inelastic scattering (SIDIS) processes, we have:
\be
d\sigma^{\ell p \to \ell' h X} = \sum_q  f_{q/p}(x, \bfk_\perp; Q^2)
\otimes d\sigma^{\ell q \to \ell q} \otimes
 D_{h/q}(z, \bfp_\perp; Q^2) \>. \label{fac}
\ee
%
\begin{figure}[pt]
\centerline{\epsfig{file=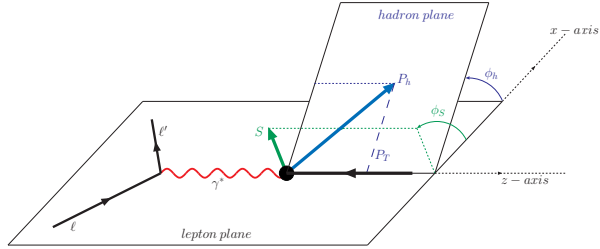,width=8cm}}
\vspace*{8pt}
\caption{A schematic illustration of SIDIS kinematics \label{SIDIS-kin}}
\end{figure}
%
In the $\gamma^* - p$ c.m.~frame, see Fig.~\ref{SIDIS-kin}, the measured transverse momentum,
$\bfP_T$, of the final hadron is generated by the transverse momentum of the
quark in the target proton, $\bfk_\perp$, and of the final hadron with respect
to the fragmenting quark, $\bfp_\perp$. At order $k_\perp/Q$ it is simply given by
\be
\bfP_T =  z \,\bfk_\perp + \bfp_\perp \>.
\ee
There is a general consensus
that such a scheme holds in the kinematical
region defined by
\be
P_T \simeq \Lambda_{\rm QCD} \ll Q \>.\label{kin}
\ee
The presence of the two scales, small $P_T$ and large $Q$, allows to
identify the contribution from the unintegrated partonic distribution
($P_T \simeq \kt$), while remaining in the region of validity of the QCD
parton model.

Within this simple scheme we can successfully describe a wide range of
unpolarized and polarized experimental data, provided we are able to model 
and phenomenologically determine the appropriate TMDs, including 
their scale evolution.
Historically, in the Torino-Cagliari standard approach, TMDs are 
parametrized in a form in which their dependence on the lightcone momentum 
fraction and on the partonic intrinsic transverse momentum are factorized, 
\bea
&&f_{q/p} (x,\kt)= f_{q/p} (x,Q^2)\,\frac{e^{-\kt^2/\avk}}{\pi\avk}\,,
\label{unp-dist}\\
&&D_{h/q}(z,\pp)=D_{h/q}(z,Q^2)\,\frac{e^{-\pp^2/\avp}}{\pi\avp}\,,
\label{unp-frag}
\eea
with a $Q^2$-independent, normalized Gaussian factor giving the intrinsic transverse 
momentum distribution, multiplied by a collinear unpolarized parton 
distribution function (PDF) evolving with $Q^2$  
according to DGLAP equations; $\avk$ and $\avp$ are free parameters which can be 
extracted from experiments.
A similar parameterization is deviced for polarized TMDs, like the Sivers function
\begin{equation}
 \Delta^N \! f_ {q/\pup}(x,\kt) = 2 \, {\cal N}_q(x) \, h(\kt) \,f_ {q/p} (x,\kt,Q^2)
 \label{siv-dist}
\end{equation}
with
\begin{equation}
f_ {q/p} (x,\kt)=f(x,Q^2)\frac{
e^{-k_\perp^2/\langle k_{\perp}^2\rangle}}
{\pi\langle k_{\perp}^2\rangle}\,,
\end{equation}
\begin{equation}
{\cal N}_q(x) =  N_q \, x^{\alpha_q}(1-x)^{\beta_q} \,
\frac{(\alpha_q+\beta_q)^{(\alpha_q+\beta_q)}}
{\alpha_q^{\alpha_q} \beta_q^{\beta_q}}\,,
\end{equation}
and
\begin{equation}
h(\kt) = \sqrt{2e}\,\frac{k_\perp}{M_{1}}\,e^{-{k_\perp^2}/{M_{1}^2}}\,.
\end{equation}
$N_q$, $\alpha_q$, $\beta_q$ and $M_1$ are free parameters 
which can be extracted from experimental data.

A completely analogous parameterization holds for the Collins function.

In Ref.~[\refcite{Anselmino:2005nn}] we performed a study of the unpolarized 
and Sivers TMDs based on EMC\cite{Aubert:1983cz} measurements of the azimuthal dependence 
and the $P_T^2$ distribution of the SIDIS cross sections, where we were able 
to extract the Gaussian widths, $\avk$ and $\avp$, appearing in 
Eqs.~(\ref{unp-dist}) and ~(\ref{unp-frag}), which represented the two free parameters 
of our fit. We found:
\be
\avk=0.25\; {\rm GeV}^2 \;, \qquad \avp=0.20\; {\rm GeV}^2 \;.
\ee

Afterwards, in Refs.~[\refcite{Anselmino:2005ea,Anselmino:2008sga,Anselmino:2010bs}] by exploiting 
the above parameters, we extracted the Sivers function by fitting polarized SIDIS azimuthal Sivers 
asymmetries from the HERMES~\cite{Airapetian:2004tw,Airapetian:2009ae} and COMPASS~\cite{Alekseev:2008aa} experiments. 
Similarly, in a combined fit~\cite{Anselmino:2007fs,Anselmino:2008jk,Anselmino:2013vqa} of 
SIDIS~\cite{Airapetian:2004tw,Airapetian:2010ds,Ageev:2006da,Adolph:2012sn} 
and $e^+e^-$~\cite{Abe:2005zx,Seidl:2008xc,Seidl:2012er} experimental data, 
we extracted the transversity and Collins functions simultaneously.
%
\begin{figure}[pb]
\centerline{\epsfig{file=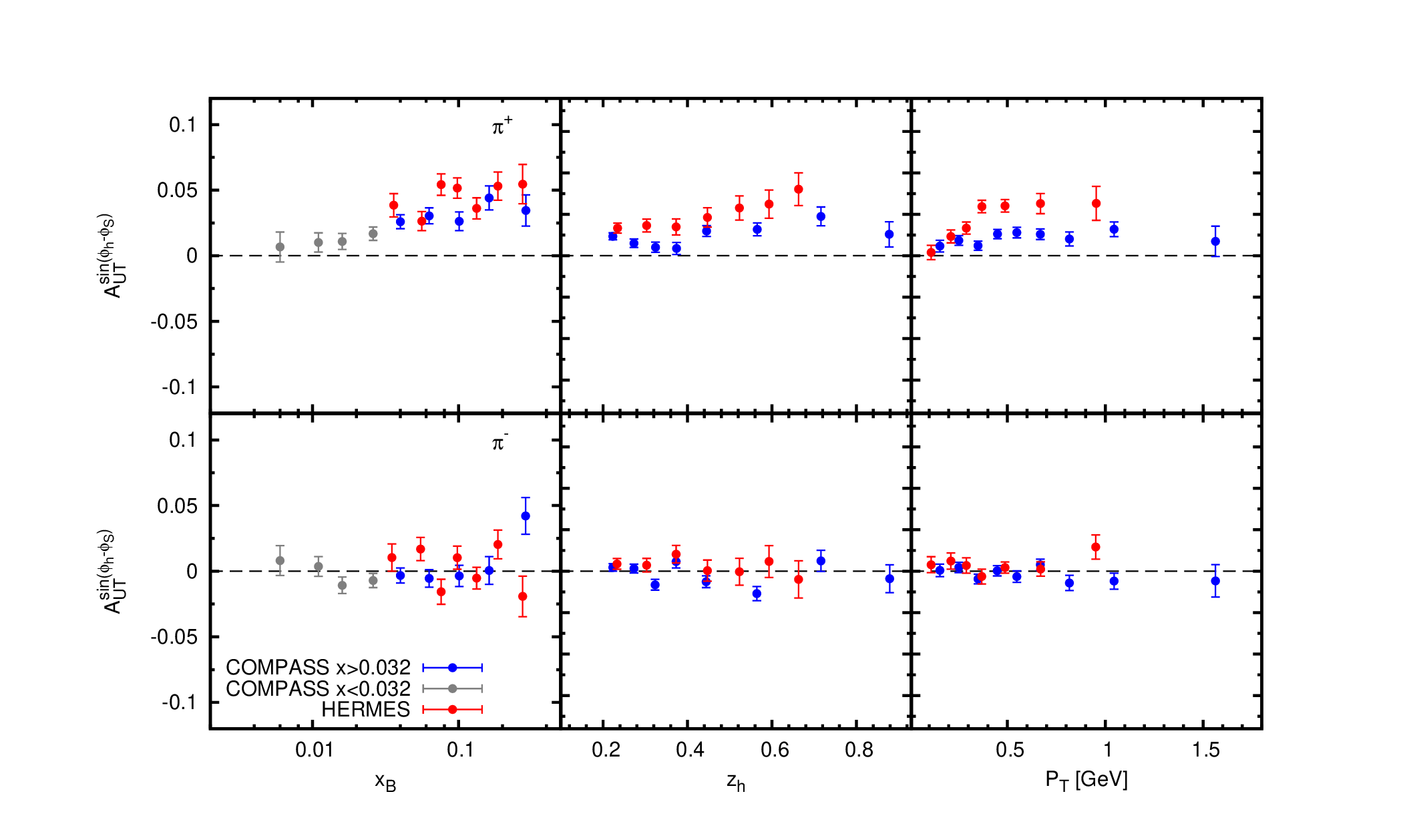,width=6.0cm,angle=-90}} 
\centerline{\epsfig{file=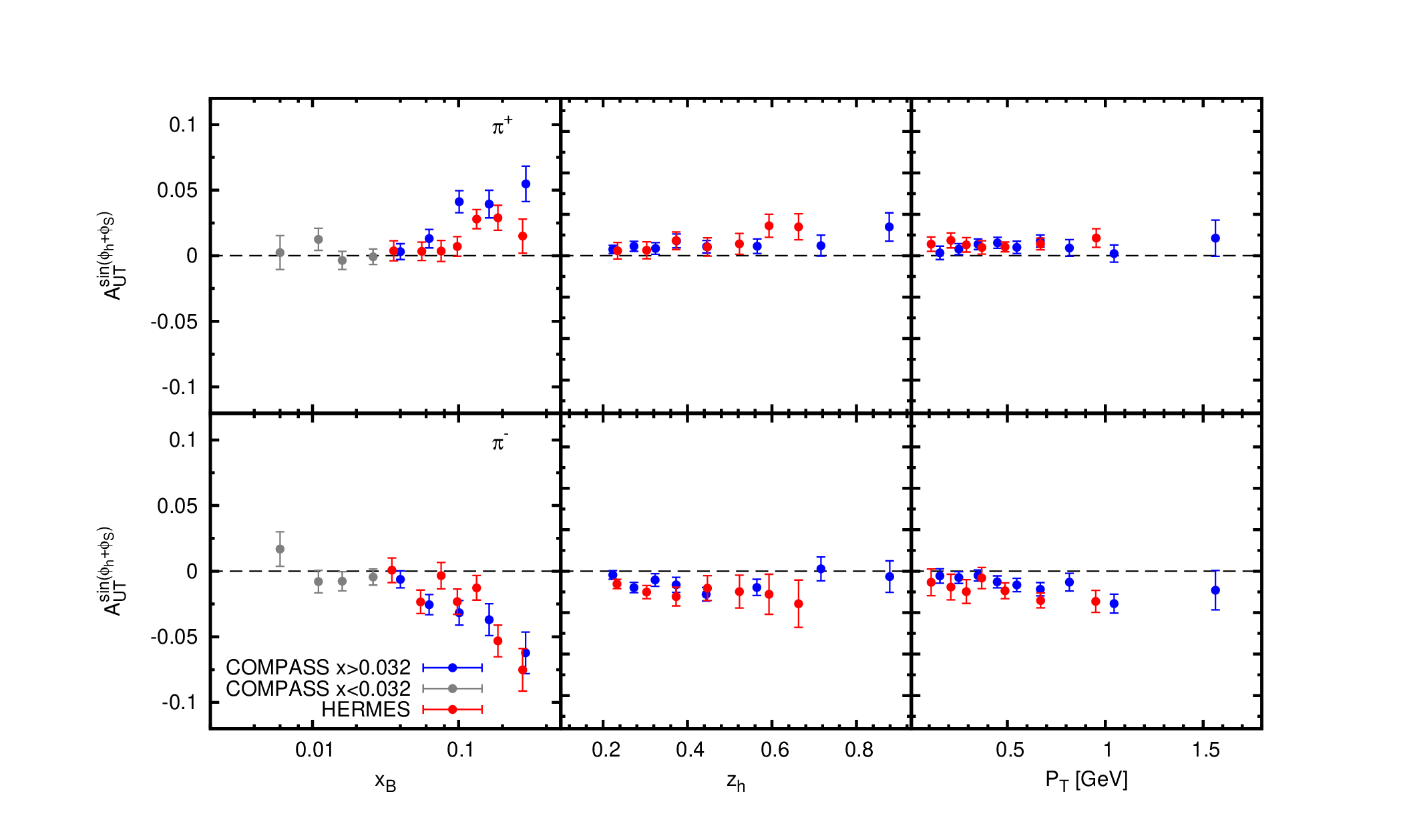,width=6.0cm,angle=-90}}
\caption{Comparison between the Sivers $A_{UT}^{\sin(\phi_h-\phi_S)}$ asymmetry, upper panel,
and Collins $A_{UT}^{\sin(\phi_h+\phi_S)}$ asymmetry, lower panel, as measured by the 
HERMES~\protect\cite{Airapetian:2009ae,Airapetian:2010ds} and 
COMPASS~\protect\cite{Adolph:2012sn,Adolph:2012sp} Collaborations. \label{HERM-COMP}}
\end{figure}

As new, more precise and higher statistics data are rapidly becoming available, it is due time  
to start wondering whether we can find any sign of TMD scale evolution in the experimental 
measurements we can now access. Interesting studies\cite{Martin:2013eja} have been performed by comparing the Sivers 
asymmetries measured by the HERMES\cite{Airapetian:2009ae,Airapetian:2010ds} and COMPASS\cite{Adolph:2012sn,Adolph:2012sp} Collaborations, over similar 
$x$, $z$ and $P_T$ ranges, but different average values of $Q^2$: 
\be
\langle Q^2 \rangle = 2.4\; {\rm GeV}^2 \;\;{\rm for ~ HERMES} \;\; {\rm and}  \;\;
\langle Q^2 \rangle = 3.2\; {\rm GeV}^2\;\; {\rm for ~ COMPASS}\,.
\ee
As shown in Fig.~\ref{HERM-COMP} it is quite evident that there is a systematic difference between 
the two sets of data, which can possibly be interpreted as a sign of scale evolution. 
A similar comparison for the analogous Collins asymmetry shows that in this case there is 
no evident inconsistency between the two measurements, and no indication of a possible TMD evolution 
for the Collins function can be deduced.

From the theory point of view, much progress has recently been achieved in the framework of TMD evolution.
After the 1985 first pioneering work on QCD resummation for transverse momentum dependent processes by Collins, 
Soper and Sterman\cite{Collins:1984kg}, the issue was revived by Idilbi, Ji, Ma and Yuan 
in several papers\cite{Ji:2004xq,Idilbi:2004vb,Ji:2004wu} in 2004--2005 and, later on, revisited 
by J. Collins in his 2011 book\cite{Collins:2011book}. This scheme of TMD evolution was finalized by 
Aybat and Rogers\cite{Aybat:2011zv} for the unpolarized TMDs, who also proposed, together with  J. Collins 
and J. Qiu, a model for the TMD evolution for polarized parton densities, in particular for the Sivers 
function\cite{Aybat:2011ge}.

Most of the people mentioned above are participants of this workshop, therefore the interested reader can find more 
details and better explanations in their contributions to these proceedings.

\begin{figure}[pt]
\centerline{\epsfig{file=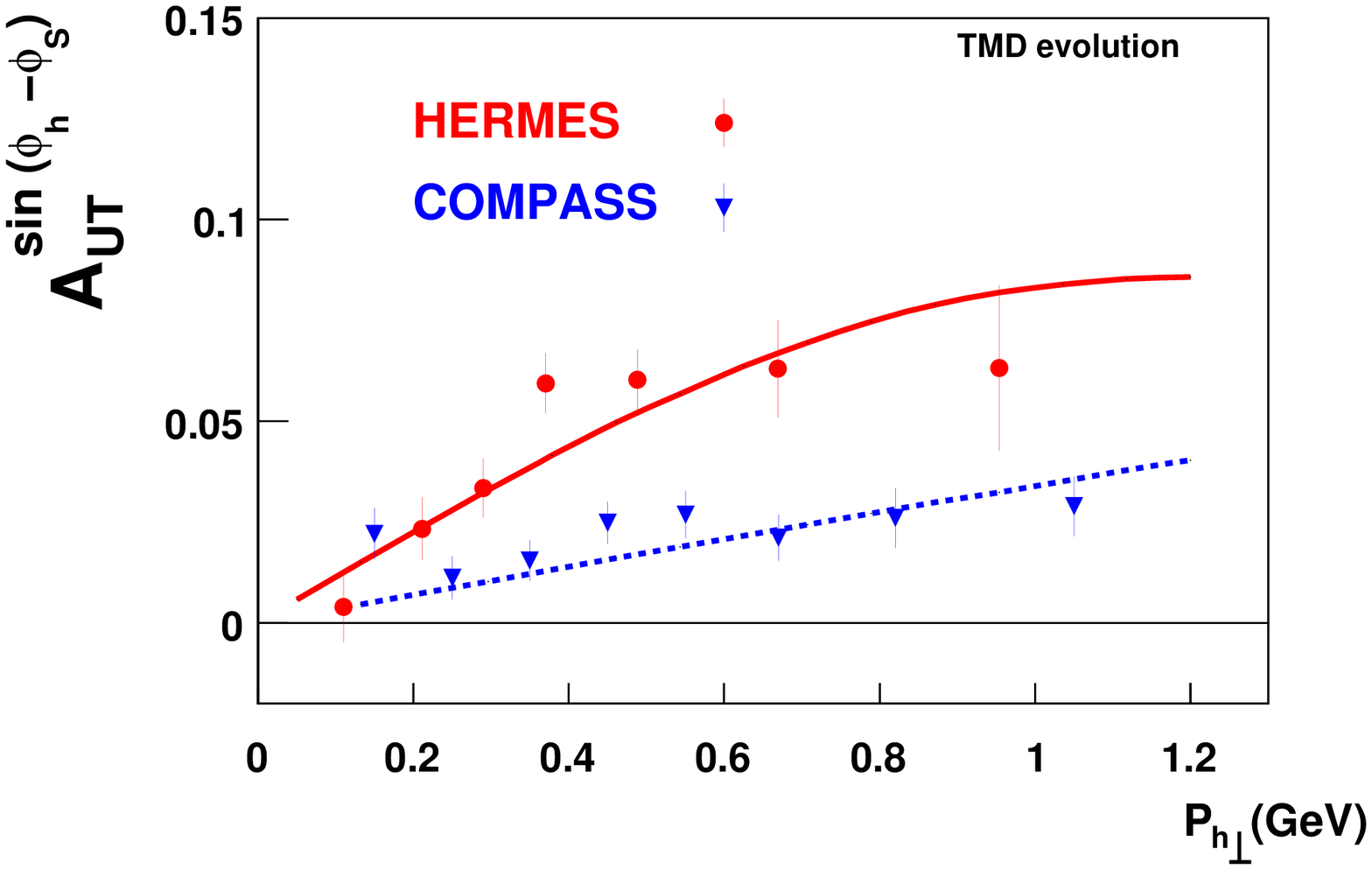,width=6.0cm} \hspace*{0.5cm} \epsfig{file=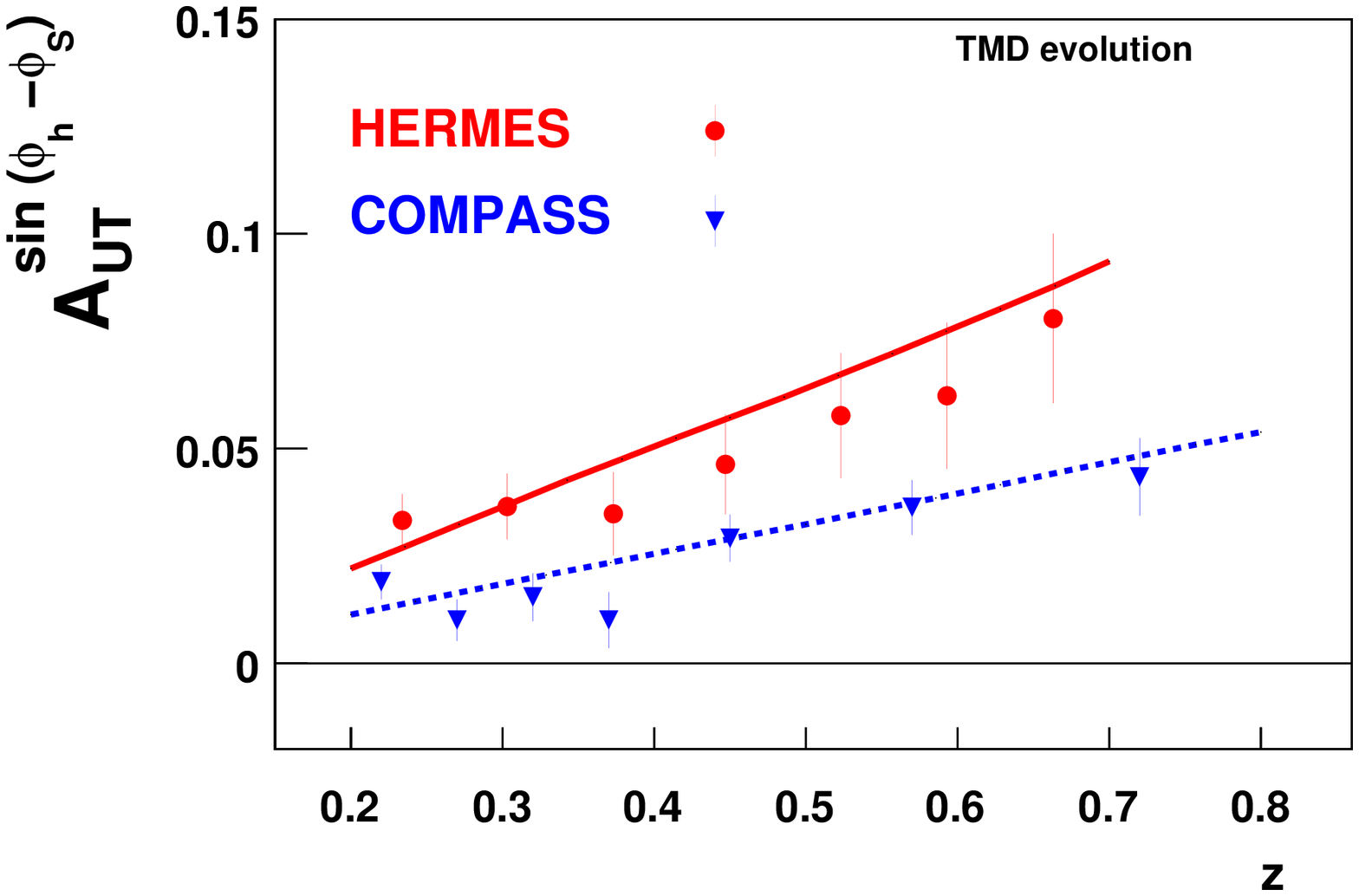,width=6.0cm}}
\caption{Comparison between HERMES\protect\cite{Airapetian:2009ae} and COMPASS\protect\cite{Adolph:2012sn} 
data and the analogous asymmetry obtained from the phenomenological model of 
Ref.~[\protect\refcite{Anselmino:2008sga}] 
by applying the TMD evolution of  Ref.~[\protect\refcite{Aybat:2011ge}] (dashed line) 
or simply using Eq.~(\protect\ref{siv-dist}) (solid line).
\label{alexei}}
\end{figure}

After the publication of these theory works, the first studies of what can more properly be defined TMD-phenomenology 
started to be performed: Aybat, Prokudin, Rogers\cite{Aybat:2011ta} proposed a very elementary phenomenological exercise 
in which they compared the HERMES and COMPASS Sivers single spin asymmetry $A_{UT}$, calculated at their two 
fixed values of $\langle Q^2\rangle$: $2.4$ GeV$^2$ for HERMES and $3.8$ for COMPASS;
evolution effects were then compared, as illustrated in Fig.~\ref{alexei}. 
Notice that no $x$ dependence was taken into account, somehow washing out the most sensitive and interesting 
information this kind of data can provide on TMD scale evolution. 
%

\begin{figure}[pt]
\centerline{\epsfig{file=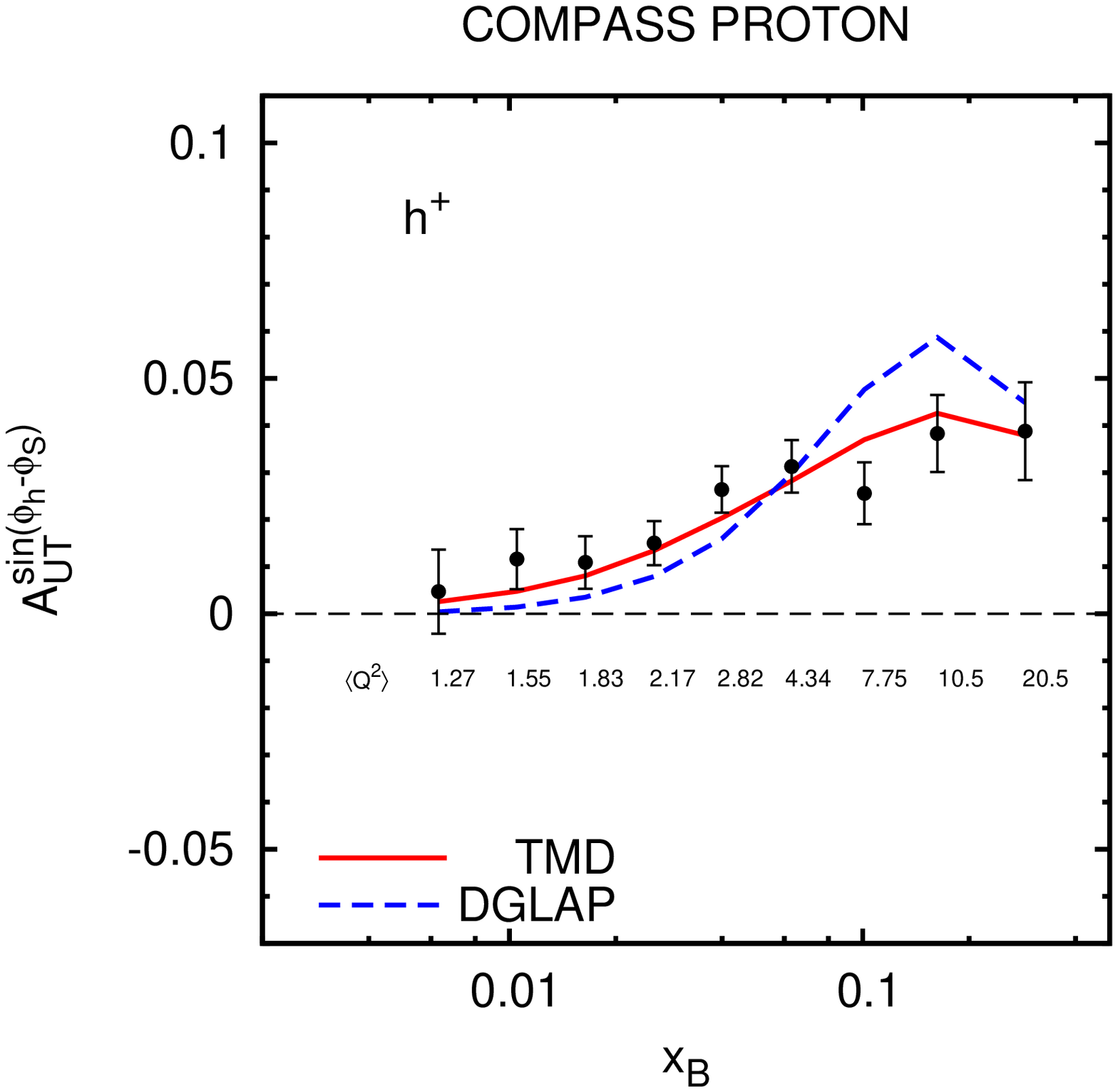,width=6.cm,angle=-90} \hspace*{-2.0cm}
\epsfig{file=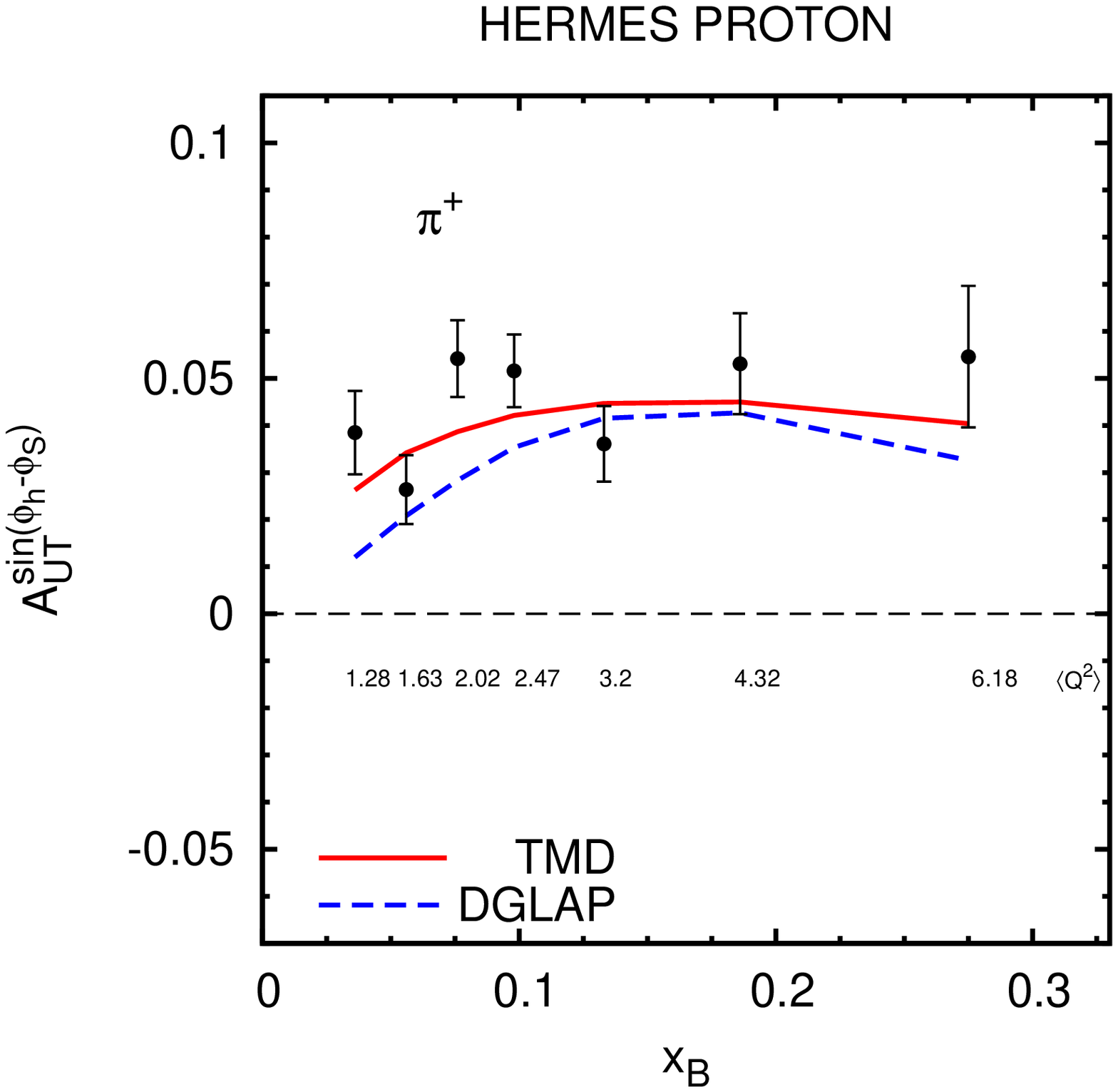,width=6.0cm,angle=-90}}
\caption{The results obtained for the SIDIS $A_{UT}^{\sin{(\phi_h-\phi_S)}}$ 
Sivers asymmetries applying TMD evolution (red, solid lines) are compared with 
the analogous results found by using DGLAP evolution equations (blue, dashed lines). 
The green, dash-dotted lines correspond to the results obtained by 
using the approximated analytical TMD evolution (see text for further details). 
The experimental data are from HERMES~\protect\cite{Airapetian:2009ae} (left panel) and 
COMPASS~\protect\cite{Adolph:2012sn} (right panel) Collaborations. 
\label{noi}}
\end{figure}

At the same time, M. Anselmino, S. Melis and myself~\cite{Anselmino:2012aa} performed a new, more refined phenomenological analysis of the Sivers 
TMD evolution, again based on the comparison between HERMES and COMPASS sets of measurements on the Sivers effect, 
but focussing mostly on the actual $Q^2$ and $x$ dependence.   We have shown that 
TMD-evolution\cite{Aybat:2011zv,Aybat:2011ge} implies a strong 
variation with $Q^2$ of the functional form of the unpolarized and Sivers 
TMDs, as functions of the intrinsic momentum $\kt$; moreover, our fit 
of all SIDIS data on the Sivers asymmetry using TMD-evolution, when compared 
with the same analysis performed with the simplified DGLAP-evolution, exhibits 
a smaller value of the total $\chi^2$, a reduction which 
mostly originates from the large $Q^2$ COMPASS data, which are greatly 
affected by the TMD evolution, as shown in Fig.~\ref{noi}. 
We then considered this as a strong indication in favor 
of the TMD evolution.  
%

Without going into the details of such a complex evolution scheme, 
there is one particular point that I would like to address.  
The TMD evolution equation of the unpolarized TMD PDFs, in 
configuration space, is the following
\be
\widetilde f_{q/p}(x, \bt; Q) =  f_{q/p}(x,Q_0) \;\widetilde R(Q, Q_0, \bt)\;
\exp \left\{-b_T^2 \left(\avk/4  + \frac{g_2}{2} \ln \frac{Q}{Q_0}\right) \right\} \,,
\label{evF-f}
\ee
\be
{\rm where}\; g_K(b_T) = \frac12 \, g_2 \, b_T^2 \;\; {\rm with} \;\;
g_2 = 0.68  \> {\rm GeV}^{2} \;\;
{\rm corresponding~to} \;\; b_{\rm max}=0.5 \> {\rm GeV}^{-1} \>.
\label{gk}
\ee
A similar equation regulates the evolution of the first derivative of the Sivers function.
Eq.~(\ref{evF-f}) shows that the $Q^2$ evolution is 
controlled by the logarithmic $Q$ dependence of the $b_T$ Gaussian width, 
together with the factor $\widetilde R(Q, Q_0, \bt)$: for increasing values 
of $Q^2$, they are responsible for the typical broadening effect already 
observed in Refs.~[\refcite{Aybat:2011zv}] and~[\refcite{Aybat:2011ge}].
Notice that the parameter $g_2$, that controls the $\bt$ Gaussian width 
and its spreading, is not extracted from our fit, but 
taken as a fixed values from elsewhere~\cite{Landry:2002ix}. We could have determined its value in our fit, 
and probably got a smaller value, but it is important to remember that SIDIS asymmetries are very little 
sensitive to the precise value of $g_2$; therefore, our choice to keep it fixed was motivated by the balance one always has 
to keep between number of data and number of free parameters, to obtain a reliable fit.
Drell Yan processes, instead, where $Q^2$s are much larger and perturbative corrections become important, 
are extremely sensitive to it. Therefore, one should not simply take the parameters used in this application 
of TMD evolution to SIDIS processes and apply them blindy to Drell Yan data or other processes. 
This would require a new, careful, global analysis on all SIDIS and Drell Yan data, re-starting 
from unpolarized cross sections. 
\begin{figure}[pt]
\centerline{\epsfig{file=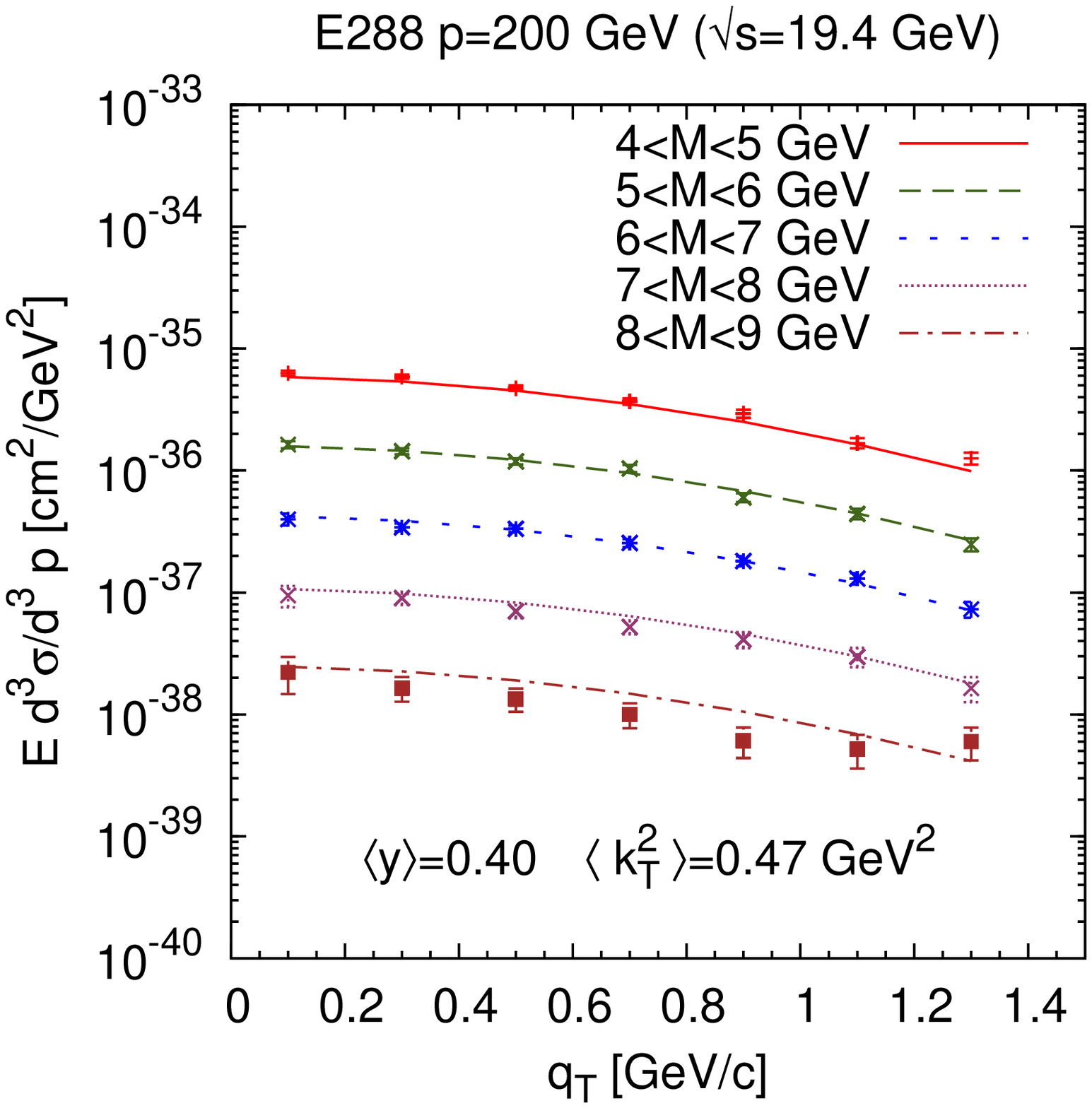,width=4.7cm,angle=-90} \hspace*{-1.1cm} 
\epsfig{file=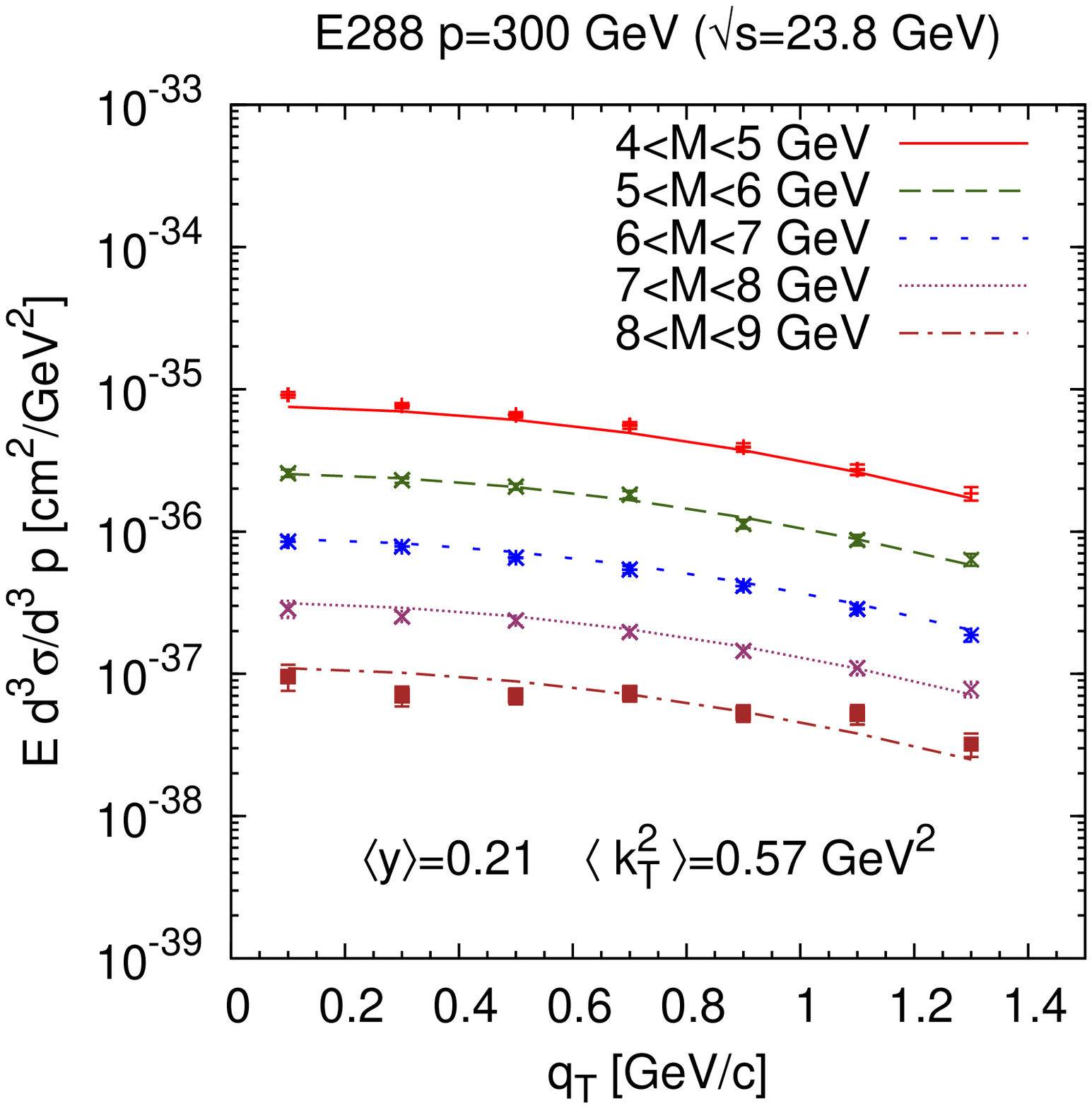,width=4.7cm,angle=-90}}
\centerline{\epsfig{file=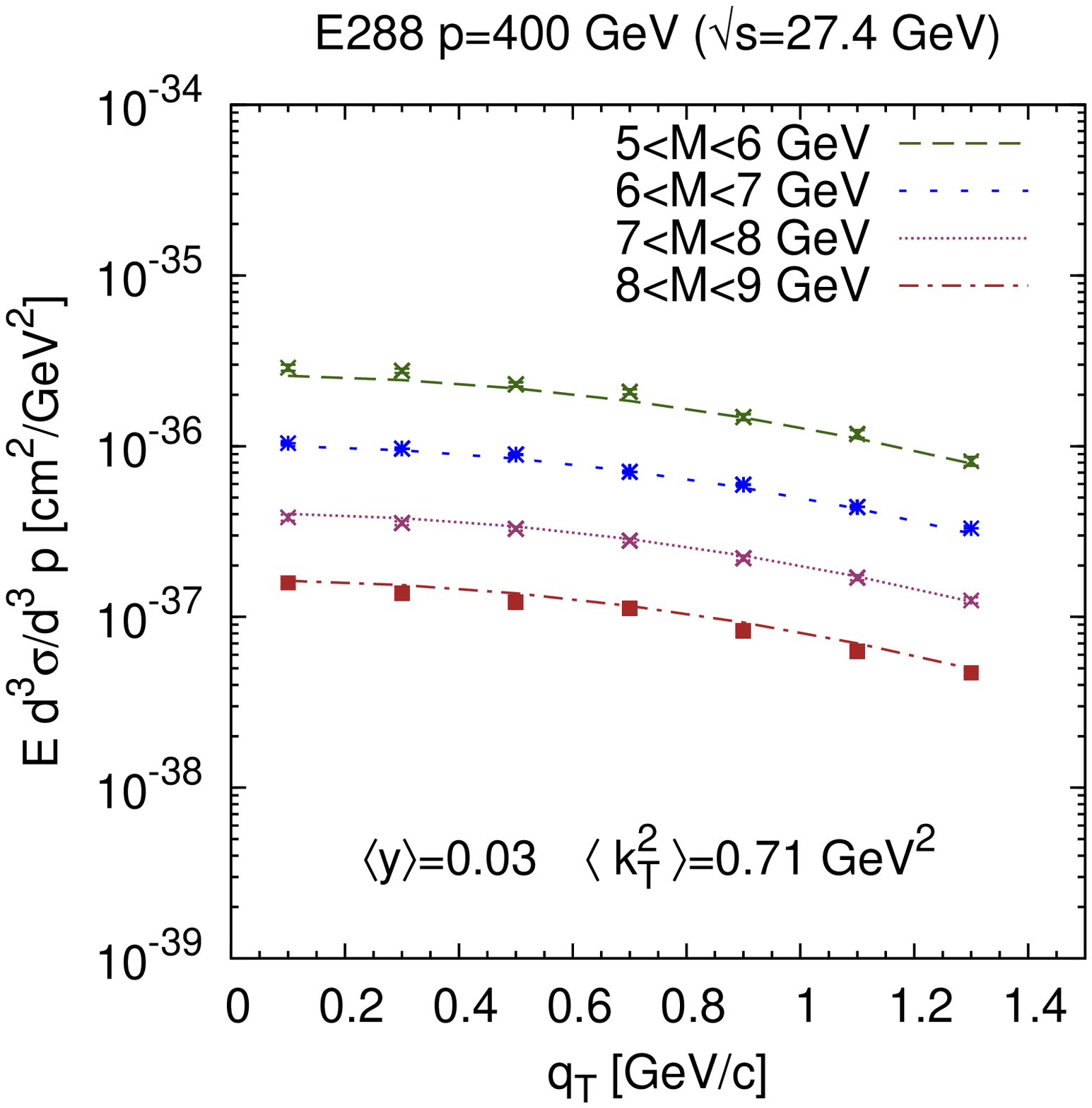,width=4.7cm,angle=-90} \hspace*{-1.1cm} 
\epsfig{file=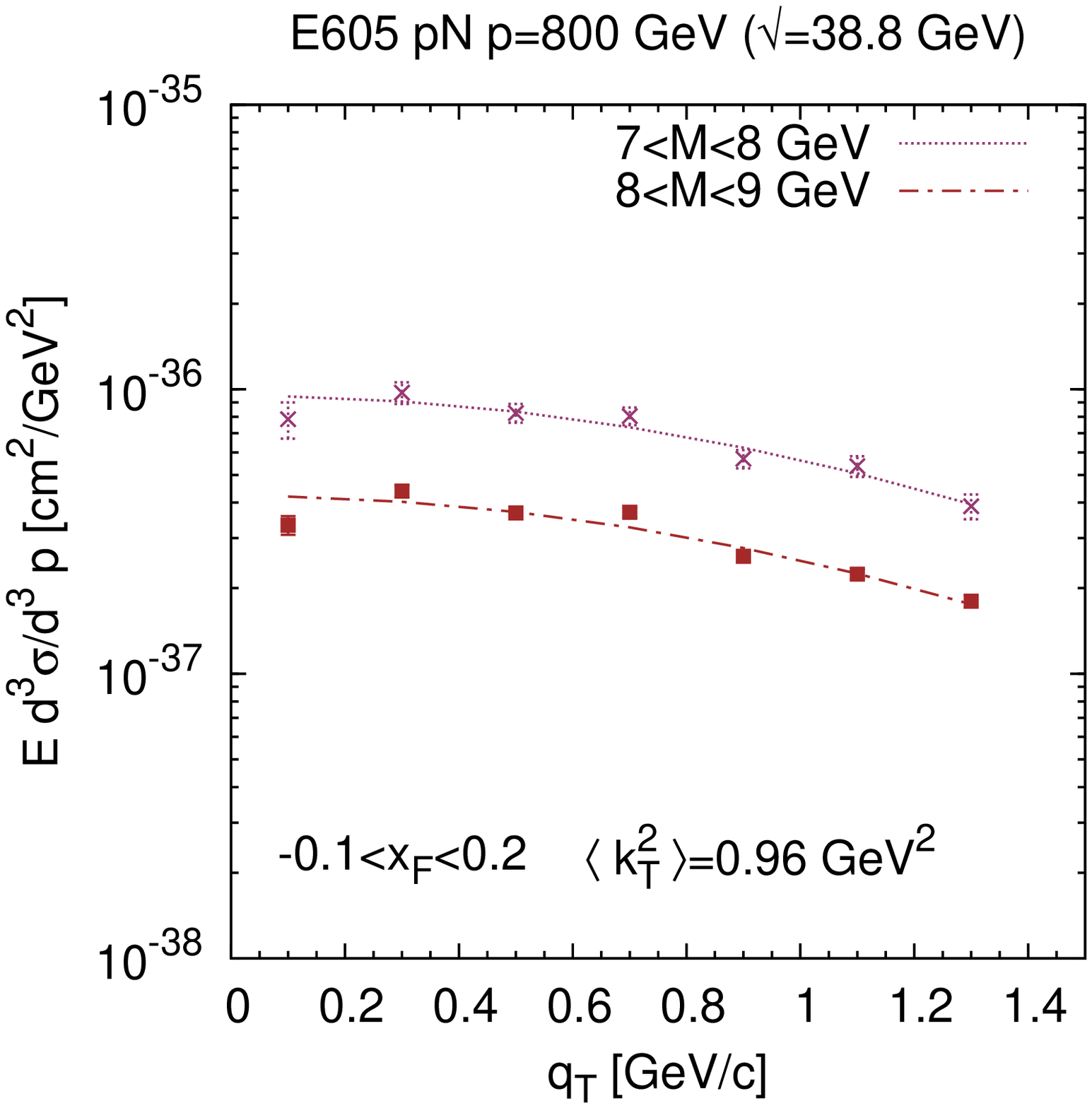,width=4.7cm,angle=-90} }
\caption{Gaussian fit of Drell Yan data from E288\protect\cite{Ito:1980ev} 
and E605\protect\cite{Moreno:1990sf} experiments.
\label{stefano1}}
\end{figure}

Sun and Yuan\cite{Sun:2013dya} have recently applied some CSS\cite{Collins:1984kg}--like evolution scheme at
one loop, with strong approximations which hold for moderate $Q$ and $Q_0$, to account
for the TMD evolution of the unpolarized TMD PDFs, and extended this formalism to the
Sivers function as well. In this approximated scheme the evolution does not produce such a strong suppression
of the Drell-Yan asymmetries. They then perform a phenomenological study on a rather limited selection 
of Drell-Yan and SIDIS data, showing that the evolution scheme they propose can satisfactorily describe most of them.

\begin{figure}[pt]
\centerline{
\epsfig{file=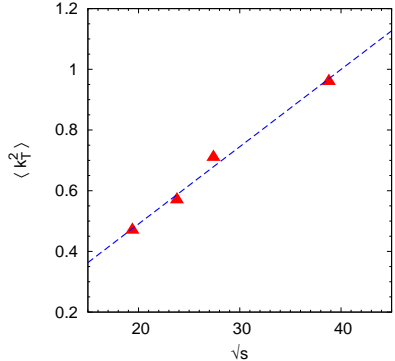,width=5.0cm,angle=-90} }
\caption{Average $\kt$ extracted from Drell Yan data as a function of $\sqrt{s}$:
notice that $\avk$ shows a linear growth with $\sqrt{s}$.
\label{stefano2}}
\end{figure}

Remaining on Drell Yan processes, a recent preliminary study was performed by S. Melis, 
fitting the E288\cite{Ito:1980ev} and E605\cite{Moreno:1990sf} Drell-Yan data and 
assuming a Gaussian $\kt$ dependence. 
The free parameter $\avk$ was extracted separately for each data sets 
corresponding to different $\sqrt{s}$ values. The results showed that those Drell Yan 
data indicate a $\sqrt{s}$ (roughly linear) depedence in addition to the logarithmic $Q^2$
dependence typical of scale evolution. This is shown in Figs.~\ref{stefano1} and~\ref{stefano2}.

Most recently D. Boer\cite{Boer:2013zca} has performed a study of the energy scale
dependence of the Sivers asymmetry in SIDIS, although on a larger range of $Q$ values
($3$ - $100$ GeV): he finds that the peak of the Sivers asymmetry falls off with $Q$ roughly like
$(1/Q)^{0.7}$ , quite faster than found within the CSS~\cite{Collins:1984kg} evolution schemes.
Moreover, the peak of the asymmetry is located around the initial scale $Q_0$ and moves rather
slowly towards higher transverse momentum values as $Q$ increases, which
may be due to the absence of perturbative tails of the TMDs.

Important work on TMD evolution has also recently been done by Echevarria, Idilbi, 
Scimemi\cite{GarciaEchevarria:2011rb}  in the framework of 
effective field theory which, however, has not yet reached the stage of feasible 
phenomenological applications.

As this is one of the opening talks of this workshop, there are no proper conclusions. 
Rather, I will close with a few remarks on future perspectives.

As far as TMD evolution is concerned, we have recently come a long way.
We now have evolution schemes and some first attempts towards a full phenomenological 
study of the unpolarized distribution and fragmentation TMDs, and of the 
Sivers and Collins functions.
These are very preliminary studies, which need to be refined and re-thought in a more 
consistent and appropriate way, especially as far as the parametrization of unknown 
phenomenological quantities are concerned.  

From the experimental side, we certainly need more SIDIS (polarized and unpolarized) 
data at larger values of $x$ (JLab 12) and spanning a larger $Q^2$ range (EIC) as well as more 
(and more precise) Drell-Yan data, for which many beautiful experiments are being planned 
(COMPASS, RHIC, Fermilab, NICA, JPARK).
Inclusive hadron production in hadron-hadron scattering processes, as well, represent a 
very interesting and infinitely challenging 
field where to “sharpen our tools”.

With the new experimental data on SIDIS multiplicities coming in, we have to go back one 
step, re-think and re-perform a solid, global analysis of Drell-Yan as well as SIDIS 
unpolarized cross sections, to determine the basic parameters for the phenomenological 
quantities needed for the implementation of the TMD evolution schemes.
Afterwards, we can proceed on a firm footing to perform the same analysis for the Sivers, 
transversity and Collins TMD functions, keeping in mind the importance of finding 
phenomenological frameworks suitable for all processes.

\section*{Acknowledgments}

I thank the organizers for their warm hospitality; in particular, Alexei 
Prokudin who, beside being a great collaborator, always manages to turn a 
conference into a very special event. I am most grateful to S. Melis and 
M. Anselmino for their relentless work and constant support.




\end{document}